\def\a{\alpha}
\def\b{\beta}
\def\m{\mu}
\def\t{\theta}
\def\d{\delta}
\def\p{\partial}
\def\L{\Lambda}

\documentstyle[12pt,epsf]{article}

\begin{document}
\begin{titlepage} 
\vspace*{1.0cm}
\begin{center} {\LARGE \bf  
Fit to Gluon Propagator and \\[0.3cm]
Gribov Formula} \\
\vspace*{0.5cm}
{\bf Attilio Cucchieri$^{\star}$}
and {\bf Daniel Zwanziger$^{\ddag}$}\\ \vspace*{0.7cm}
${}^{\star}$IFSC-USP, Caixa Postal 369, 
             13560-970 S\~ao Carlos, SP, Brazil \\
\vspace*{0.4cm}
${}^{\ddag}$Physics Department, New York University,
	     New York, NY 10003, USA
\\ \vspace*{1.4cm}
{\bf   Abstract   \\ } \end{center} \indent
We report a numerical study of $SU(2)$ lattice
gauge theory in the minimal Coulomb
gauge at $\b = 2.2$ and 9 different volumes.
After extrapolation to infinite volume, our fit
agrees with a lattice discretization of Gribov's
formula for the transverse
equal-time would-be physical gluon propagator,
that {\it vanishes}
like $|{\bf k}|$ at ${\bf k} = \bf{0}$, whereas
the free equal-time propagator $(2|{\bf k}|)^{-1}$ {\it
diverges}.
Our fit lends reality to a
confinement scenario in which the would-be physical gluons leave the
physical spectrum while the long-range Coulomb force confines color.
\vfill \begin{flushleft} 
\noindent{\rule[-.3cm]{5cm}{.02cm}} \\
\vspace*{0.2cm} \hspace*{0.5cm}
\noindent
Electronic addresses: {\tt attilio@if.sc.usp.br}
and {\tt daniel.zwanziger@nyu.edu}
\end{flushleft} 
\end{titlepage}


\section{Introduction}

	In QCD a rectangular Wilson loop $W(R, T)$ of dimension $R \times T$
has, asymptotically at large $T$, the form $W(R,T) \sim \exp[-T V_W(R)]$,
where $V_W(R)$ is the Wilson potential.  If dynamical quarks are
present, they are polarized from the vacuum, and $V_W(R)$ represents the
interaction energy of a pair of mesons at separation $R$.  In
this case $V_W(R)$ is not a color-confining potential, but rather
a QCD analog of the van der Waals potential between neutral atoms.  It
clearly cannot serve as an order parameter for confinement of color in
the presence of dynamical quarks, and we turn instead to gauge-dependent
quantities to characterize color confinement.

\begin{figure}
\begin{center}
\vspace*{-5.0cm} \hspace*{0cm}
\epsfxsize=0.90\textwidth
\leavevmode\epsffile{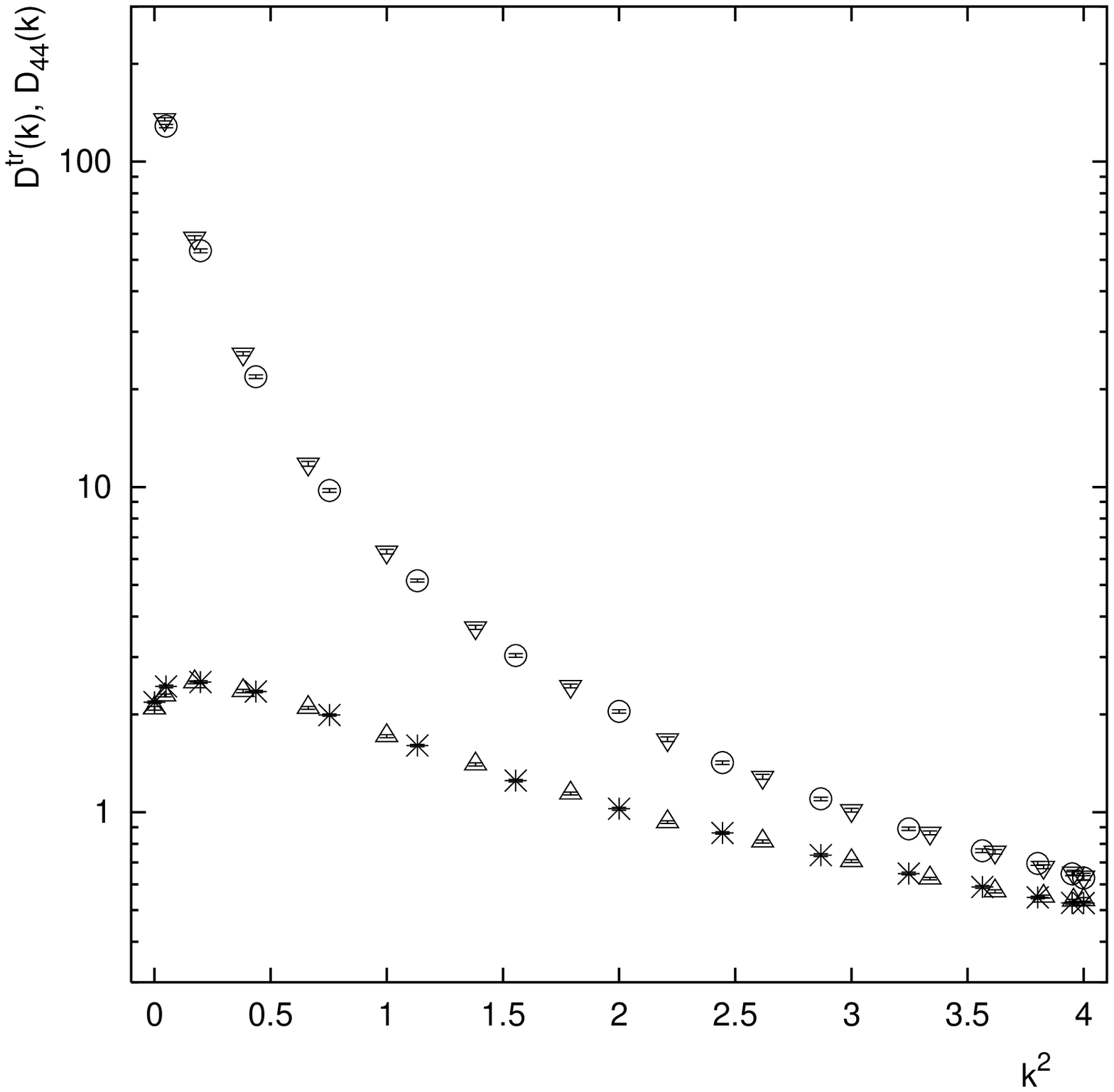}\\
\caption{~Plot of the gluon propagators
$D^{\rm tr}({\bf k})$ and $D_{44}({\bf k})$ as a function
of the square of the lattice momentum ${\bf k}^2$ for
$L = 28$ (symbols $\ast$ and $\bigcirc$ respectively) and
$L = 30$ (symbols $\triangle$ and $\bigtriangledown$ respectively).
Notice the logarithmic scale in the $y$ axis.
Error bars are one standard deviation.}
\label{FIG:compar}
\end{center}
\end{figure}

	A particularly simple confinement scenario \cite{gribov,coul} is
available in the minimal Coulomb gauge.\footnote{The minimal lattice
Coulomb gauge is obtained by first minimizing
$-\sum_{x,i} {\rm Tr} {^gU}_{x,i}$ with respect to all local gauge
transformations $g(x)$, and then minimizing
$-\sum_x {\rm Tr} {^gU}_{x,4}$ with respect 
to all ${\bf x}$-independent but $x_4$-dependent 
gauge transformations $g(x_4)$. This makes the
3-vector potential $A_i$, for $i = 1,2,3$ transverse,
$\p_i A_i = 0$, so $A_i = A_i^{\rm tr}$.
Moreover, the Coulomb gauge is the finite limit of renormalizable
gauges \cite{BZ}.
\label{Foot:minimal}}
It attributes confinement
of color to the {\it enhancement} at long range of the color-Coulomb
potential $V(R)$.  This is the instantaneous part of the 4-4
component of the gluon propagator,
$D_{44}({\bf x}, t) = V(|{\bf x}|) \d(t) + P({\bf x}, t)$, where
the vacuum polarization term $P({\bf x}, t)$ is less singular at
$t = 0$.  At the same time, the disappearance of gluons from the
physical spectrum is manifested by the {\it suppression} at ${\bf k} = \bf{0}$
of the propagator $D_{ij}({\bf k}, k_4)$ of 3-dimensionally transverse
would-be physical gluons.  This behavior is clearly exhibited in
Fig.\ \ref{FIG:compar}
which displays the equal-time propagators
$D^{\rm tr}({\bf k})$ and $D_{44}({\bf k})$.

	We conjecture that the color-Coulomb potential $V(R)$ is linearly
rising
at large $R$ (at least when asymptotic freedom holds) and that this
linear rise may serve as an order parameter for color confinement in the
presence or absence of dynamical quarks \cite{coul}.
As a practical matter,
$V(R)$ is the starting point for calculations of bound states such as
heavy quarkonium
\cite{V}.  Remarkably, $V(R)$ is a
renormalization-group invariant \cite{coul} in the sense that it is
independent of the cut-off $\L$ and of the renormalization mass $\m$.
Its Fourier transform $\widetilde{V}({\bf k})$ may serve to define the
running coupling constant of QCD by
${\bf k}^2\widetilde{V}({\bf k})
= x_0 \ g_{\rm coul}^2(|{\bf k}|)$.
Here $x_0 = 12 N/(11N - 2N_f)$ was calculated in \cite{rgcoul}, and
$g_{\rm coul}(|{\bf k}|)$ satisfies the perturbative renormalization
group equation
$|{\bf k}| \p g_{\rm coul} / \p |{\bf k}|
= \b_{\rm coul}(g_{\rm coul})$.  It has the leading asymptotic behavior
$g_{\rm coul}^2(|{\bf k}|)
\sim [2b_0 \ln(|{\bf k}|/\L_{\rm coul})]^{-1}$,
where
$b_0 = (4 \pi)^{-2} (11N - 2N_f) / 3 $, and
$\L_{\rm coul} \propto \L_{\rm QCD}$ is a finite QCD mass scale.
These formulas allow one to determine the running coupling
constant of QCD from a numerical evaluation of the equal-time 2-point
function $D_{44}$ in the minimal Coulomb gauge.  By  contrast, in
Lorentz-covariant gauges, the coupling constant $g_r(\m)$ requires a
numerical evaluation of the 3-point vertex function \cite{gluonalf}.
Numerical studies of the gluon propagator in Landau gauge may be found
in \cite{mandula,langfeld}.

A less intuitive but equally striking prediction concerns the
3-dimensionally transverse, would-be physical gluon propagator
$D_{ij}({\bf k}, k_4) =
(\d_{ij} - \hat{k}_i\hat{k}_j)D^{\rm tr}({\bf k}, k_4)$,
whose equal-time part is given by
$D^{\rm tr}({\bf k}) = (2\pi)^{-1} \int dk_4 D^{\rm tr}({\bf k}, k_4)$.
It was proven \cite{vanish} as a consequence of the Gribov horizon
that, for infinite spatial lattice volume $L^3$, $D^{\rm tr}({\bf k})$
{\it vanishes} at ${\bf k} = \bf{0}$,
$\lim_{|{\bf k}| \to 0}D^{\rm tr}({\bf k}) = 0$,
although the rate of approach of
$D^{\rm tr}({\bf k}, L)$ to 0, as a function of  ${\bf k}$ or
$L$, was not established, nor was it determined whether the renormalized
gluon propagator also shares this property.  This is in marked
contrast to the free massless propagator $({\bf k}^2 + k_4^2)^{-1}$
which at equal-time is given by
$(2\pi)^{-1} \int dk_4 ({\bf k}^2 + k_4^2)^{-1} = (2|{\bf k}|)^{-1}$, that
{\it diverges} at ${\bf k} = \bf{0}$.  Gribov \cite{gribov} obtained, for the
energy of a gluon of momentum ${\bf k}$, the approximate expression
$E({\bf k}) = |{\bf k}|^{-1}[({\bf k}^2)^2 + M^4]^{1/2}$.  This gives for
the equal-time, 3-dimensionally transverse, would-be physical equal-time
gluon propagator the approximate formula \footnote{There are
corresponding results for the 
Landau gauge \cite{gribov,vanish,vanish2}.
Also in Landau gauge, recent solutions of the Schwinger-Dyson for 
the gluon propagator $D(k)$ vanish at $k = 0$ \cite{smekal}.}
\begin{equation}
D^{\rm tr}({\bf k}) = [2E({\bf k})]^{-1}
= (|{\bf k}|/2)[({\bf k}^2)^2 + M^4]^{-1/2} \, .
\label{approp}
\end{equation}
In the absence of an estimate of corrections, one does not know how
accurate this formula may be.  Its accuracy, and the crucial
question of the extrapolation to large $L$ of $D^{\rm tr}({\bf k}, L)$ will
be addressed in the fits reported here.

	We wish to confront Gribov's theory of confinement in the minimal
Coulomb gauge \cite{gribov,coul} with data from our
numerical study of $SU(2)$ lattice gauge theory, without quarks, in the
minimal Coulomb
gauge at $\beta = 2.2\, $.
Data were taken at 9 different lattice
volumes $L^4$, with
$L = 14, 16, 18, \ldots, 30$
in order to extrapolate to infinite lattice volume.
The lattice Coulomb gauge is more easily accessible to numerical study
than the Landau gauge because each time-slice contributes separately to
the numerical average which,  for a lattice of volume $30^4$, gives a
factor of 30 gain.
Details of the numerical simulation are described in
\cite{cuzwns}. (For this study we have however improved the statistics
for the lattice volume $18^4$.) The minimal Coulomb gauge is
obtained by an over-relaxation algorithm which, in general, leads to some
one of the relative minima of the minimizing functional (see
footnote \ref{Foot:minimal}),
each of which is a different Gribov copy. As discussed in the Appendix
of \cite{cuzwns}, the gluon propagator in Coulomb gauge is insensitive,
within statistical uncertainty, as to which Gribov copy is attained, in
agreement with similar results in the Landau gauge \cite{langfeld,cumen}. 

We reported fits of $\widetilde{V}({\bf k})$ and $D^{\rm tr}({\bf k}, L)$ in
\cite{cuzwns}.  Our data clearly show that $\widetilde{V}({\bf k})$ is more
singular than $1/{\bf k}^2$ at low ${\bf k}$, which indeed corresponds to
a long-range color-Coulomb potential.  [A linearly
rising potential $V(R)$ corresponds to
$\widetilde{V}({\bf k}) \sim 1/({\bf k}^2)^2$]. However,
an extrapolation in $\b$ will be necessary to
determine the strength of this singularity in the continuum limit,
because $U_4$ is quite far from the continuum limit at $\b = 2.2$, as
compared to $U_i$, for $i = 1,2,3$, due to the gauge-fixing
in the minimal Coulomb gauge that is described in footnote \ref{Foot:minimal}.
In the present Letter we present a new fit to $D^{\rm tr}({\bf k}, L)$,
and most importantly, its extrapolation to infinite lattice volume $L \to
\infty$, using a lattice discretization of Gribov's formula.

\section{Fit to $D^{\rm tr}({\bf k})$}

To parametrize the data we choose a fitting formula which: (i) gives a good
fit to $D^{\rm tr}({\bf k}, L)$
for all momenta ${\bf k}$ and volumes $L^4$, (ii) includes Gribov's
formula (\ref{approp}) as a special case, and (iii) is physically transparent.
Recall that a free continuum field with
mass $m$ has the equal-time propagator
$(2\pi)^{-1} \int dk_4 \ (k_4^2 + {\bf k}^2 + m^2)^{-1}
= [2({\bf k}^2 + m^2)^{1/2}]^{-1}$.
With this in mind, consider the continuum
formula \footnote{In Ref.\ \cite{cuzwns} we made a fit to a formula
corresponding to a sum of 2 poles. However,
it did not include the Gribov formula as a special case.}
\begin{equation}
D^{\rm tr}({\bf k}) = 4^{-1} [z \ ({\bf k}^2)^\a + s]
 [({\bf k}^2 + m_1^2)({\bf k}^2 + m_2^2)]^{-1/2}
\label{contf}
\end{equation}
with singularities corresponding to poles at $m_1^2$ and $m_2^2$, which
we take to be either a pair of real numbers or a complex-conjugate
pair.\footnote{According to the general principles of 
quantum field theory, the propagator of
physical particles should have poles only at real
positive $m^2$. However, in the confined 
phase the gluon propagator may have singularities
that correspond to unphysical excitations.}
For complex-conjugate poles $m_1^2 = m_2^{*2} = x + iy$, this reads
\begin{equation}
D^{\rm tr}({\bf k}) = 4^{-1} [z \ ({\bf k}^2)^\a + s]
[({\bf k}^2 + x)^2 + y^2)]^{-1/2} \, .
\label{contfa}
\end{equation}
The case of a pair of real poles is obtained from this formula by taking
$y^2 < 0$.  Gribov's formula (\ref{approp}) is recovered for
$\a = 0.5,\  s = 0$, $x = 0$, and $y = M^2$ (and $z = 2$),
corresponding to a pair of purely imaginary poles.

	To obtain a lattice discretization of (\ref{contf}), note that for
$k_\m = 2\sin(\t_\m/2)$, the lattice free propagator at equal time is
given by
$(2\pi)^{-1} \int_0^{2\pi} d\t_4 (k_4^2 + {\bf k}^2 + m_1^2)^{-1}
= (4h_1 + h_1^2)^{-1/2}$, where $h_1 \equiv {\bf k}^2 + m_1^2$.  This
suggests discretizing (\ref{contf}) by the substitution,
$4h_1 \to 4h_1 + h_1^2$, and similarly for $1 \to 2$.
For $m_1^2 = m_2^{*2} = x + iy$, one has
$4h_1 + h_1^2 = u + 2iyv$, where
$u \equiv 4({\bf k}^2 + x) + ({\bf k}^2 + x)^2 - y^2$, and
$v \equiv (2 + {\bf k}^2 +x)$.  This gives the lattice fitting formula that
we shall use,
\begin{equation}
D^{\rm tr}({\bf k}) =
[z \ ({\bf k}^2)^\a + s] (u^2 + 4y^2v^2)^{-1/2} \, ,
\label{mod}
\end{equation}
where $k_i = 2\sin(\t_i/2)$, and
$ -\pi \leq \theta_i = 2 \pi n_i/L \leq \pi$, for integer $n_i$.
In the continuum limit one has $u \to 4({\bf k}^2 + x)$ and $v \to 2$, and
(\ref{mod}) approaches (\ref{contfa}).

\begin{figure}
\begin{center}
\vspace*{-5.0cm} \hspace*{0cm}
\epsfxsize=0.90\textwidth
\leavevmode\epsffile{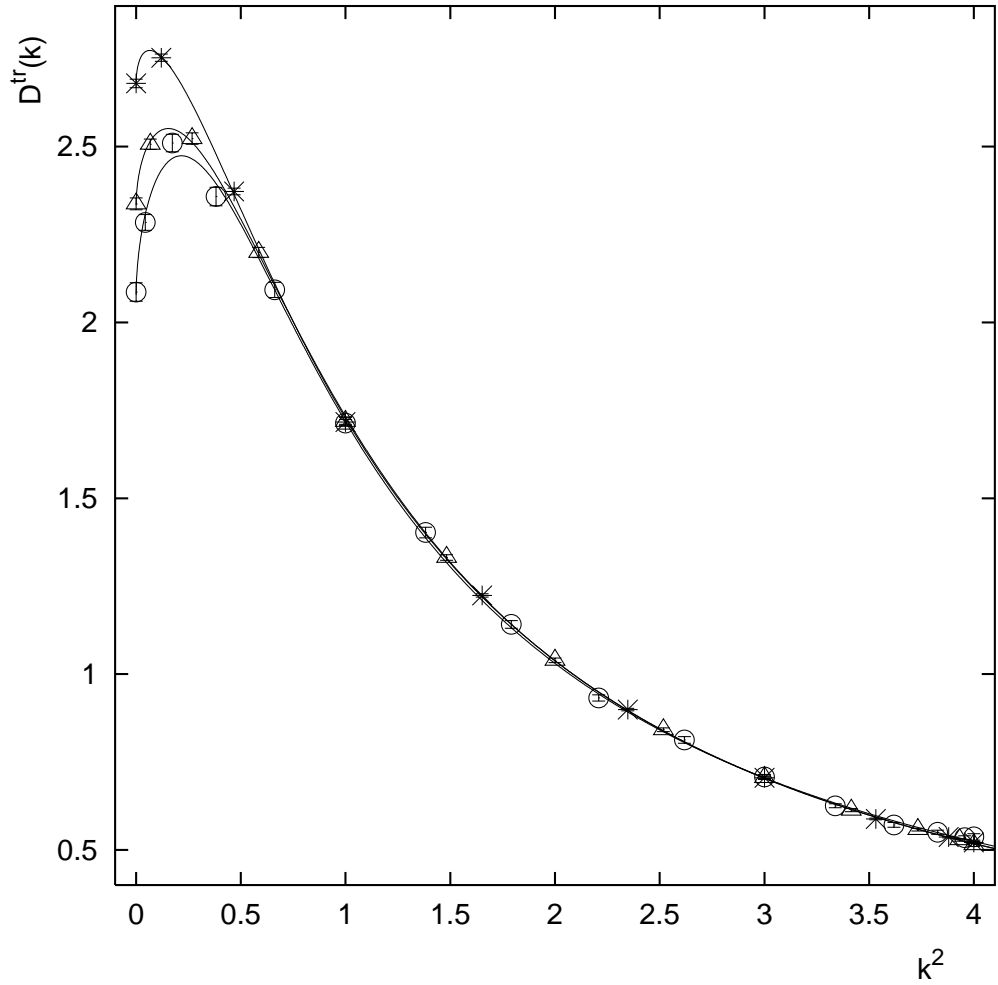}\\
\caption{~Plot of the gluon propagator $D^{\rm tr}({\bf k})$
as a function
of the square of the lattice momentum ${\bf k}^2$
for $L = 18 (\ast), \, 24 (\triangle)$ and $30 (\bigcirc)$ and fits of
these data using Eq.\ (\ref{mod}).
Error bars are one standard deviation.}
\label{FIG:fitt}
\end{center}
\end{figure}

	For each lattice size $L$ we have made a
fit of the parameters $z(L),\, s(L),\, \a(L),\, x(L)$ and $y^2(L)$
for the nine lattice volumes considered.
Fig.\ \ref{FIG:fitt}
shows our fit for $D^{\rm tr}({\bf k})$ for $L = 18, 24, 30$, and
the fit to the other volumes are comparable.
There is no a priori reason why a 2-pole fit should be
accurate over the whole range of momenta considered.\footnote{In
our simulations we consider only 3-momenta 
aligned along major axes $\theta_i = (0, 0,
2\pi n/L)$. Thus, the maximum momentum value
(for each lattice side $L$) is $2$ in lattice 
units, and $1.876$ GeV in physical units (with 
the inverse lattice spacing $a^{-1}$ set equal 
to $0.938$ GeV, see Ref.\ \cite{cuzwns}).}
However, the fit is excellent for all momenta ${\bf k}^2$ and
for each $L$. A striking feature of the fit is that $y^2(L)$ is positive
for all 9 values of $L$, corresponding to a pair of
{\it complex}-conjugate poles rather than a pair of real
poles. 
Complex poles are also seen
in the minimal Landau gauge \cite{landaupoles}
and for gauges that interpolate between them \cite{gluonhotnew},
and at finite temperature \cite{gluonhotnew,gluonhot}.
By contrast, in the maximal
Abelian gauge \cite{gluonhotnew,gluonhot,maxabel} and in
Laplacian gauge \cite{deforcrand}, poles are observed to occur at real
$k^2$.

\begin{figure}
\begin{center}
\vspace*{-5.0cm} \hspace*{0cm}
\epsfxsize=0.90\textwidth
\leavevmode\epsffile{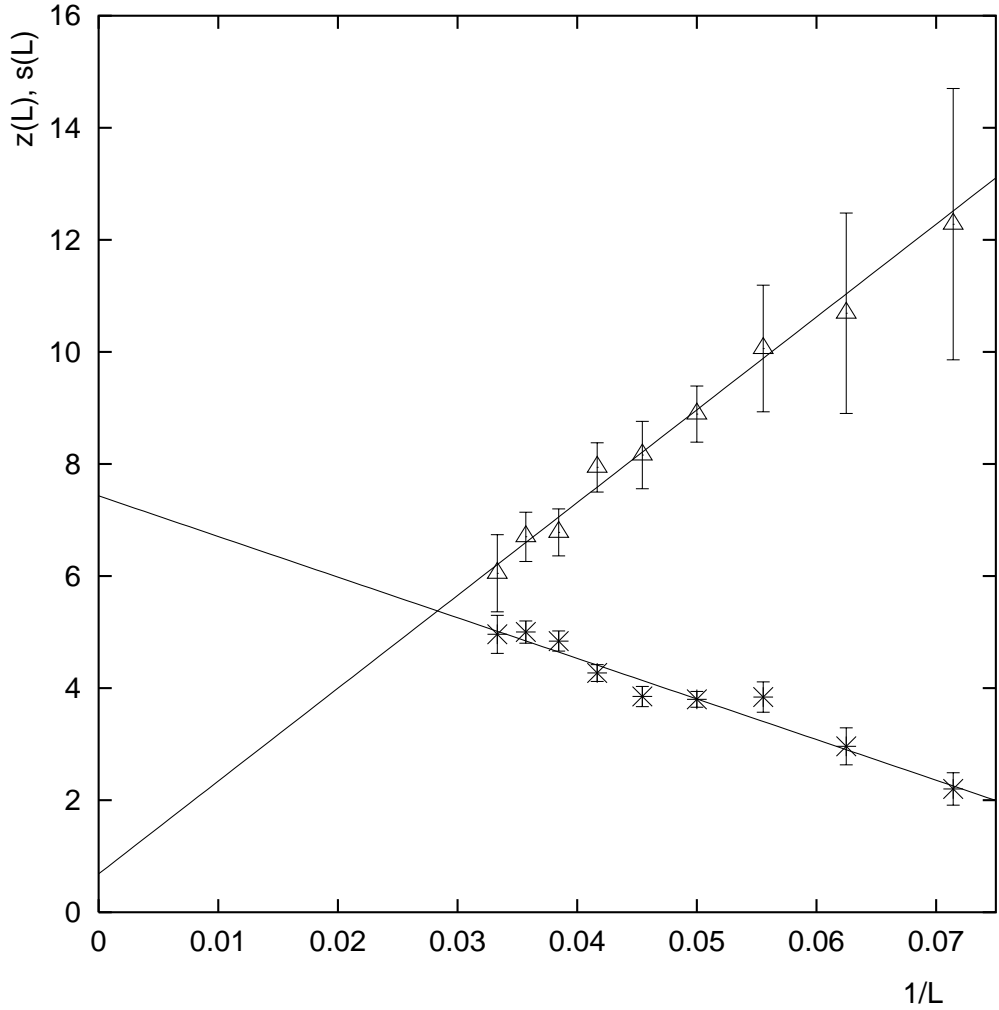} \\
\caption{~Fits of the parameters $z(L) (\ast)$ and $s(L) (\triangle)$
plotted against $1/L$.
For $z(L) = a + b/L$, we obtain
$a = 7.43 \pm 0.36$,
$b = -72.5 \pm 7.7$, and $\chi^2/$d.o.f\ $= 1.13$.
For $s(L) = a + b/L$, we obtain
$a = 0.685 \pm 0.534$,
$b = 165.6 \pm 12.7$, and $\chi^2/$d.o.f\ $=  0.183$.}
\label{FIG:zz}
\end{center}
\end{figure}

\begin{figure}
\begin{center}
\vspace*{-5.0cm} \hspace*{0cm}
\epsfxsize=0.90\textwidth
\leavevmode\epsffile{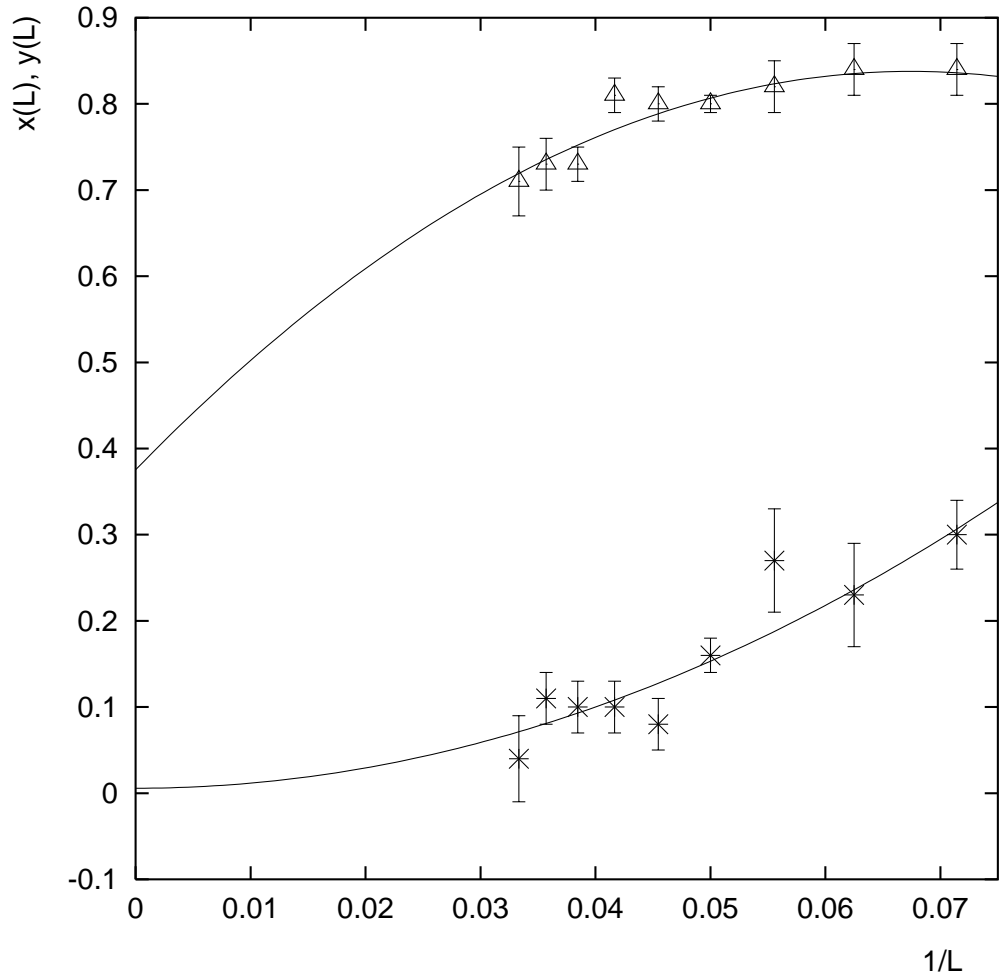}\\
\caption{~Fits of the parameters $x(L) (\ast)$ and $y(L) (\triangle)$
plotted against $1/L$. For $x(L) = a + c/L^2$
we obtain
$a = 0.006 \pm 0.024$,
$c = 59.0  \pm 9.9$, and
$\chi^2/$d.o.f.\ $= 0.858$.
For $y(L) = a + b/L + c/L^2$
we obtain
$a = 0.375 \pm 0.162$,
$b = 13.7  \pm 6.3$,
$c = -101.6  \pm 61.0$, and
$\chi^2/$d.o.f.\ $= 1.024$.}
\label{FIG:xx}
\end{center}
\end{figure}

\begin{figure}
\begin{center}
\vspace*{-5.0cm} \hspace*{0cm}
\epsfxsize=0.90\textwidth
\leavevmode\epsffile{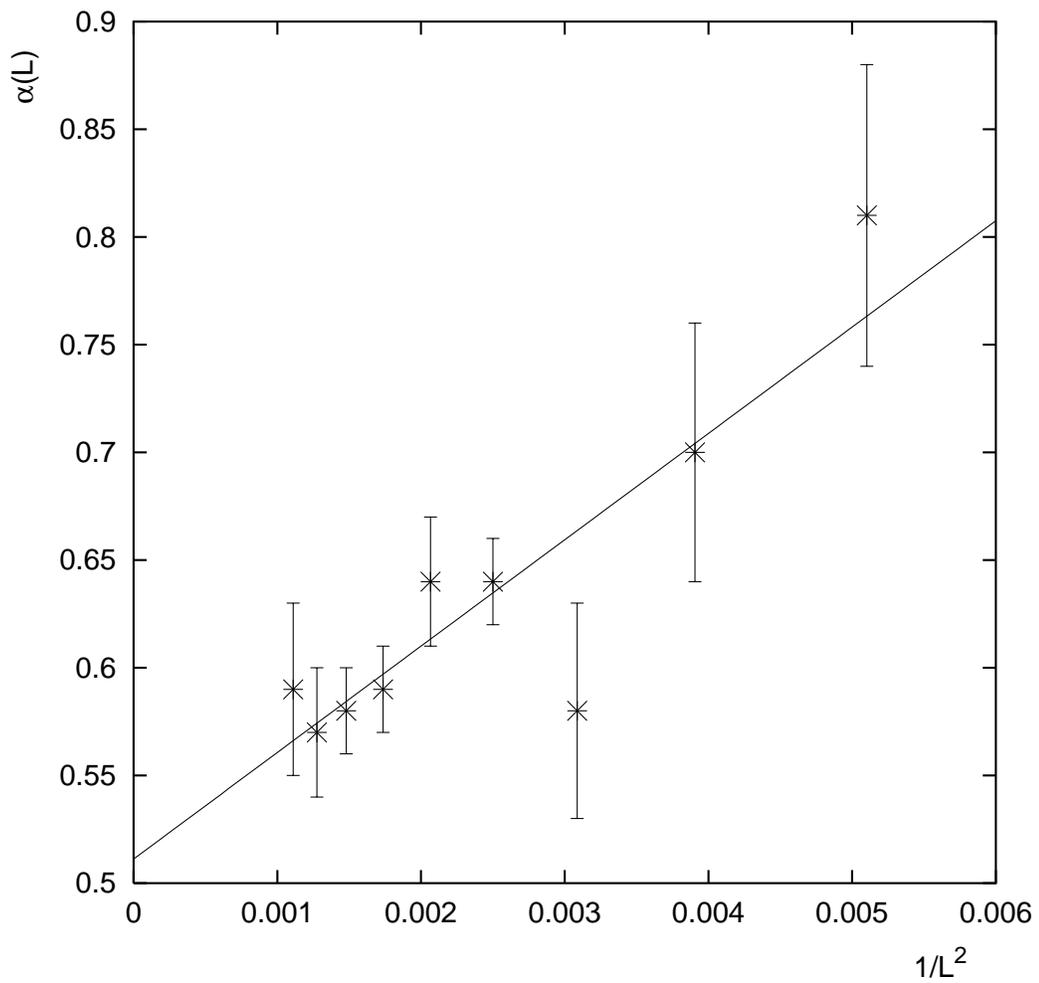}\\
\caption{~Fit of the parameter $\alpha(L)$ plotted
against $1/L^2$.  For $\a = a + c/L^2$, we obtain
$a = 0.511 \pm 0.022$, $c = 49.4  \pm  10.5$, and
$\chi^2/$d.o.f\ $= 0.666$.}
\label{FIG:aa}
\end{center}
\end{figure}

	In order to do the crucial extrapolation to
infinite lattice volume, we fit 5 curves\footnote{To facilitate
comparison with $x(L)$, we plot $y(L) > 0$.}
$z(L),\, s(L),\, \alpha(L),\, x(L)$ and $y(L)$
to our values of these fitting parameters.
After some experimentation, we used the 3 fitting
functions $a + b/L$, $a + b/L + c/L^2$, and $a + c/L^2$,
for each of the 5 parameters.  Of these 3 fits, we have
displayed, for each parameter, the one with the smallest $\chi^2$ per
degree of freedom, with values reported in the captions of
Figs.\ \ref{FIG:zz}--\ref{FIG:aa}.

\section{Conclusions}

1.\ One sees clearly from Fig.\ \ref{FIG:fitt} that the equal-time would-be
physical gluon propagator $D^{\rm tr}({\bf k}, L)$ passes through a maximum
and {\it decreases} as the momentum ${\bf k}^2$ approaches $\bf{0}$ (for fixed
$L$).  The decrease is more pronounced as the size $L$ of the lattice
increases.  This counter-intuitive behavior is direct evidence of the
suppression of low-momentum components of configurations $A^{\rm
tr}({\bf k})$ caused by the restriction to the Gribov region.

2.\ In Fig.\ 3, the extrapolated value of $s(\infty) = 0.685 \pm 0.534$
is sufficiently close to 0 (compared to $b/L \sim 6$ to 12)
that we feel justified in concluding that the gluon propagator
$D^{\rm tr}({\bf k})$ 
extrapolates to a value compatible with {\it zero} at ${\bf k} = \bf{0}$.

3.\  As a result, at infinite volume our fitting formula
behaves like
$D^{\rm tr}({\bf k}) \propto ({\bf k}^2)^\a$,
at low momentum.  The extrapolated value of $\a(L)$
at infinite volume, $\a = 0.511 \pm 0.022$,
is in striking numerical agreement with Gribov's formula
(\ref{approp}), which gives
$\a = 0.5$.

4.\  We have obtained an excellent 2-pole fit for
$D^{\rm tr}({\bf k}, L)$. Our fit
indicates that the poles occur at {\it complex}
$ m^2 = x(L) \, \pm \, i y(L)$.  The
real part $x(L)$ extrapolates to
$ 0.006  \pm 0.024$
at $L = \infty$.  Remarkably, a pair of poles in ${\bf k}^2$ at purely
imaginary $m^2 = 0\, \pm\, i y$ agrees with the Gribov equal-time
propagator
$D^{\rm tr}({\bf k}) =
(2|{\bf k}|)^{-1}[1 + M^4({\bf k}^2)^{-2}]^{-1/2}$,
with $y^2 = M^4$.  Moreover at large ${\bf k}$,
this gives a leading correction to the free
equal-time propagator of relative order $({\bf k}^2)^{-2}$ with
coefficient of dimension (mass)$^4$.  It may not be a coincidence that
this is the dimension of the gluon condensate
$\langle F^2 \rangle$, which is the lowest dimensional condensate in
QCD.  Because of the theoretical suggestiveness of our
result, we are encouraged to report the values
$m^2 = 0 \, \pm \, i y$, for
$y = 0.375 \pm 0.162$ in lattice units, or
$y = 0.330 \pm 0.142$~GeV$^2$,
$M = y^{1/2} = 0.575 \pm 0.124$~GeV for the location of the gluon
poles in~${\bf k}^2$. Here it is assumed that we are already in
the scaling region for $D^{\rm tr}({\bf k})$, which remains to be
verified by further numerical studies. That this is not entirely
unreasonable is suggested by the fact that scaling for the Landau gauge
propagator in $SU(2)$ has been observed \cite{langfeld}
in the range $2.1 \leq \beta \leq 2.6$. Note also that Gribov 
derived his formula in the
continuum theory, namely at infinite $\beta$, so on the theoretical side
it is to be expected that our fit will remain valid throughout the
scaling region.

5.\   The observed strong {\it enhancement} \cite{cuzwns}
of the instantaneous
color-Coulomb potential $\widetilde{V}({\bf k})$ and the observed
{\it suppression} of the
equal-time would-be physical gluon propagator $D^{\rm tr}({\bf k})$ both at
low ${\bf k}$, strongly
support the confinement scenario of Gribov \cite{gribov,coul}.  In addition
to this qualitative agreement, we note excellent numerical
agreement of our fit to the formula
$D^{\rm tr}({\bf k}) = z(|{\bf k}|/2)[({\bf k}^2)^2 + M^4]^{-1/2}$.
Although this formula cannot be exact, it appears
that deviations from it are relatively weak at both high- and low-momentum
regimes.  If this excellent fit is maintained at larger $\b$
values, then it appears that we have obtained a quantitative
understanding of
$D^{\rm tr}({\bf k})$.

\section{Acknowledgments}
The research of Attilio Cucchieri was partially supported by
the TMR network Finite Temperature Phase Transitions in
Particle Physics, EU contract No.\ ERBFMRX-CT97-0122
and by FAPESP, Brazil (Project No.\ 00/05047-5).
The research of Daniel Zwanziger was partially supported by the National
Science Foundation under grant PHY-0099393.

\clearpage


\end{document}